%% file: main.tex
\documentclass[ALICE,manyauthors]{cernphprep}
\usepackage[comma,square,numbers,sort&compress]{natbib}
\usepackage{hyperref}
\usepackage{lineno}
\usepackage{xspace}
\usepackage{xcolor}
\usepackage[T1]{fontenc}
\usepackage{orcidlink}

\newcommand{ \la }{\langle}
\newcommand{ \ra }{\rangle}

\def\PbPb  {\mbox{Pb--Pb}\xspace}
\def\pPb   {\mbox{p--Pb}\xspace}

\def\snn   {\mbox{$\sqrt{s_{_{\rm NN}}}$}\xspace}
\newcommand{\snnPb}{\mbox{$\sqrt{s_{_{\rm NN}}} = 5.02~\mathrm{TeV}$}\xspace}
\newcommand{\snnPbOne}{\mbox{$\sqrt{s_{_{\rm NN}}} = 2.76~\mathrm{TeV}$}\xspace}
\newcommand{\cent}[2]{\mbox{$#1\text{--}#2\%$}\xspace}
\newcommand{\pt}{\ensuremath{p_{\rm T}}\xspace}
\newcommand{\pta}{\ensuremath{p_{\rm T}^{\mathrm{a}}}\xspace}
\newcommand{\ptt}{\ensuremath{p_{\rm T}^{\mathrm{t}}}\xspace}
\newcommand{\vn}{\ensuremath{v_{n}}\xspace}
\newcommand{\vncr}{\ensuremath{v_{n}\{2\}   }\xspace}
\newcommand{\Vn}[1]{\ensuremath{\vec{V}_{#1}}\xspace}
\newcommand{\PsiN}{\ensuremath{\Psi_{n}}\xspace}

\newcommand{\vnsqbr}{\ensuremath{\vn[2]}\xspace}
\newcommand{\Fpsi}[1]{\ensuremath{A_{#1}^{\rm f}}}
\newcommand{\Fvn}[1]{\ensuremath{M_{#1}^{\rm f}}}
\newcommand{\trento}{\ensuremath{\rm T_{R}ENTo}\xspace}

\graphicspath{{.img/}}

\begin{document}

\begin{titlepage}
\PHyear{2024}       
\PHnumber{084}      
\PHdate{19 March}  

\title{Systematic study of flow vector fluctuations in $\mathbf{\sqrt{\textit{s}_{_{\bf NN}}}=5.02}$ TeV $\PbPb$ collisions}
\ShortTitle{}   

\Collaboration{ALICE Collaboration\thanks{See Appendix~\ref{app:collab} for the list of collaboration members}}
\ShortAuthor{ALICE Collaboration} 

\begin{abstract}
Measurements of the \pt-dependent flow vector fluctuations in \PbPb collisions at $\snnPb$ using azimuthal correlations with the ALICE experiment at the Large Hadron Collider are presented. A four-particle correlation approach~\cite{ALICE:2022dtx} is used to quantify the effects of flow angle and magnitude fluctuations separately. This paper extends previous studies to additional centrality intervals and provides measurements of the \pt-dependent flow vector fluctuations at $\snnPb$ with two-particle correlations. Significant \pt-dependent fluctuations of the $\Vn{2}$ flow vector in Pb–Pb collisions are found across different centrality ranges, with the largest fluctuations of up to $\sim$15\% being present in the 5\% most central collisions. In parallel, no evidence of significant \pt-dependent fluctuations of $\Vn{3}$ or $\Vn{4}$ is found. Additionally, evidence of flow angle and magnitude fluctuations is observed with more than $5\sigma$ significance in central collisions. These observations in $\PbPb$ collisions indicate where the classical picture of hydrodynamic modeling with a common symmetry plane breaks down. This has implications for hard probes at high $\pt$, which might be biased by $\pt$-dependent flow angle fluctuations of at least 23\% in central collisions. Given the presented results, existing theoretical models should be re-examined to improve our understanding of initial conditions, quark--gluon plasma properties, and the dynamic evolution of the created system.
\end{abstract}

\end{titlepage}

\setcounter{page}{2} 



\section{Introduction}
Studies of ultrarelativistic heavy-ion collisions at the Relativistic Heavy Ion Collider (RHIC) and the Large Hadron Collider (LHC) have demonstrated the formation of a strongly-interacting matter called quark--gluon plasma (QGP)~\cite{Shuryak:1978ij, Shuryak:1980tp,ALICE:2022wpn,Adcox:2004mh,Arsene:2004fa,Back:2004je,Adams:2005dq}. The space--time evolution of the QGP is well described by relativistic viscous hydrodynamic models~\cite{Gale:2013da, Heinz:2013th}. An observable consequence of the QGP creation in these collisions is the anisotropic flow in the plane transverse to the beam direction~\cite{Ollitrault:1992bk, PHENIX:2003qra, STAR:2003wqp, STAR:2003xyj, ALICE:2010suc, ATLAS:2011ah, CMS:2012zex}. This anisotropy can be quantified by the Fourier decomposition of the distribution of the azimuthal angle of the final-state particles relative to the common symmetry planes~\cite{Voloshin:1994mz}
\begin{align}
	\frac{\mathrm{d}^3N}{\mathrm{d}\pt\mathrm{d}\eta\mathrm{d}\varphi}= \frac{\mathrm{d}^2N}{2\pi\mathrm{d}\pt\mathrm{d}\eta}
	\left(1+2\sum_{n=1}^\infty v_n(\pt,\eta)\cos[n(\varphi-\Psi_n(\pt,\eta))]\right),
    \label{eq:Fourier}
\end{align}
where $\varphi$ is the azimuthal angle of the emitted particles. The $v_n(\pt,\eta)$ and $\Psi_n(\pt,\eta)$ are the magnitude and orientation of the $n^{th}$-order flow vector $\Vn{n}(\pt,\eta) = v_n(\pt,\eta)e^{in\Psi_n(\pt,\eta)}$, respectively, which may depend on the transverse momentum (\pt) and the pseudorapidity ($\eta$) of the particles. This flow vector is affected by the initial collision geometry, which is dominated by the shape of the overlap region in the transverse plane between the colliding nuclei~\cite{Niemi:2012aj}. The initial anisotropy, the magnitude and orientation of which is quantified by the eccentricities $\epsilon_n$ and corresponding participant planes $\Phi_n$~\cite{Alver:2010gr, Voloshin:2007pc, Alver:2006wh}, respectively, is converted into final state momentum anisotropy by the interactions among the constituents of the QGP. For a uniform nuclear matter distribution in the initial state, the various symmetry plane angles, $\Psi_n$, coincide with the reaction plane defined by the impact parameter and beam direction for $n\geq1$~\cite{Voloshin:1994mz}. However, due to event-by-event fluctuations of the position of the nucleons inside the nuclei and of the partonic constituents inside the nucleon, the symmetry plane angles, $\Psi_n$, fluctuate around the reaction plane, leading to non-zero odd flow coefficients~\cite{Alver:2010gr, Alver:2010dn, Teaney:2010vd, Luzum:2010sp}. 
Non-zero and large values of flow coefficients have been observed at both RHIC~\cite{Arsene:2004fa, Back:2004je, Adams:2005dq, Adcox:2004mh} and the LHC~\cite{ ALICE:2010suc, ALICE:2011ab, Abelev:2014pua, Adam:2016izf, Acharya:2017zfg, ATLAS:2012at, ATLAS:2011ah, Aad:2013xma, Chatrchyan:2012wg, CMS:2012zex, Chatrchyan:2012xq}. The flow coefficients $v_n$ and their event-by-event fluctuations serve as excellent probes for constraining the initial state of heavy-ion collisions and for quantifying some of the QGP properties, such as the transport coefficients~\cite{Luzum:2012wu, Teaney:2010vd, Gale:2012rq, Niemi:2012aj, Qiu:2012uy, Teaney:2013dta, Bernhard:2019bmu, Everett:2020xug, Nijs:2020roc}.

The anisotropic flow coefficients can be measured in a $\pt$-differential way by assuming that the two-particle correlation factorizes into a product of two single-particle flow coefficients, each a function of the properties of only one of the particles. Keeping the terminology from dihadron correlation measurements of $V_{n\Delta}$~\cite{CMS:2010ifv, Aamodt:2011by, CMS:2011cqy, Abelev:2012ola}, one particle is denoted as the associated (a) and the other particle is denoted the trigger (t). The associated and trigger particles are chosen from a variable and fixed $\pt$ range, denoted $\pta$ and $\ptt$, respectively. Factorization of the two-particle correlation $V_{n\Delta}$ between the trigger and associated particles can be described as
\begin{align}
    V_{n\Delta}(\pta,\ptt) = v_n(\pta)v_n(\ptt),
    \label{eq:Facto}
\end{align}
where the $\vn(\pta)$ and $\vn(\ptt)$ are the flow coefficients for the associated and trigger particles with transverse momenta $\pta$ and $\ptt$, respectively. The factorization breaks down in hydrodynamic calculations due to the event-by-event fluctuations of the initial energy density of the heavy-ion collision~\cite{Heinz:2013bua, Gardim:2012im}. The breakdown of factorization has been observed at the LHC in \PbPb collisions at $\snn = 2.76$ TeV and \pPb collisions at $\snn = 5.02$ TeV~\cite{Acharya:2017ino, Khachatryan:2015oea, Aamodt:2011by}. This breakdown is directly related to the flow vector fluctuations in different kinematic regions. The flow vector may fluctuate as a function of \pt in both magnitude and angle~\cite{Alver:2010gr, Alver:2006wh}. As such, the flow angle, $\Psi_n(\pt)$, will 'wander' around the common symmetry plane angle $\Psi_n$~\cite{Heinz:2013bua}. This, in turn, implies that the \pt-integrated flow magnitude, $v_n$, should be interpreted as the flow of particles with respect to an integrated symmetry plane determined with particles from a specific and typically wide \pt range. The $\pt$-dependent flow angles also contribute to breaking the factorization in Eq. \eqref{eq:Facto}, as the equality in Eq. \eqref{eq:Facto} assumes a single common symmetry plane angle for all particles in an event. 

The \pt-dependent flow fluctuations can be probed with the Principal Component Analysis (PCA)~\cite{Bhalerao:2014mua, Mazeliauskas:2015vea, Mazeliauskas:2015efa, Milosevic:2018wer, Bozek:2017thv}, which can isolate subleading flow modes. The PCA has been successfully used to measure the event-by-event flow fluctuations~\cite{CMS:2017mzx} and the factorization breaking of two-particle correlations $V_{n\Delta}$ as a function of both \pt and pseudorapidity $\eta$~\cite{Hippert:2019swu}. The decorrelation effects measured in $\eta$ provide insight into the longitudinal hydrodynamic evolution of the system created in heavy-ion collisions and go beyond the assumption of a boost-invariant system used in many theoretical models~\cite{Khachatryan:2015oea, ALICE:2023tvh}. However, measurements with the PCA technique have yet to isolate the flow angle and flow magnitude fluctuations. In this paper, the flow vector fluctuations are integrated over pseudorapidity and are only studied as a function of $\pt$.

The usual way of measuring the flow vector fluctuations is with observables constructed from two-particle correlations~\cite{Heinz:2013bua, Gardim:2012im}. Such measurements have shown significant flow vector fluctuations in central $\PbPb$ collisions~\cite{Acharya:2017ino, Khachatryan:2015oea}. However, such measurements do not allow quantifying the individual contributions of the flow angle and magnitude fluctuations to the total flow vector fluctuations. Hydrodynamic models have predicted that the fluctuations of $\PsiN$ constitute more than half of the overall flow vector fluctuations~\cite{Heinz:2013bua}. Observables constructed from four-particle correlations are necessary to cancel out contributions from the flow angle or magnitude. Such observables were first presented in Ref.~\cite{ALICE:2022dtx} in selected centrality intervals and revealed significant flow angle and magnitude fluctuations in central $\PbPb$ collisions. In this paper, the study in Ref.~\cite{ALICE:2022dtx} is extended to additional centrality intervals, and measurements of flow vector fluctuations with two-particle correlations at a collision energy of $\snn = 5.02$ TeV are also presented. 

The paper is structured as follows: Section~\ref{sec:method} describes the method used to calculate the observables, while the experiment and data are described in Section~\ref{sec:exp}. The treatment of statistical and systematic uncertainties is covered in Section~\ref{sec:unc} and the results are presented in Section~\ref{sec:res}. Finally, a summary is given in Section~\ref{sec:sum}.

\section{Method}
\label{sec:method}
\noindent
The $m$-particle correlations are calculated with the Generic Framework~\cite{Bilandzic:2013kga}, an algorithm that calculates multi-particle azimuthal correlations corrected for non-uniform azimuthal detector acceptance and non-uniform detector efficiency. The flow coefficients are defined from the Fourier expansion in Eq. \eqref{eq:Fourier},
\begin{equation}
    \la v_n\ra = \la\la \cos n\left(\varphi-\Psi_n\right)\ra\ra,
\end{equation}
where the single set of brackets, $\la\ra$, denotes an average over events, while the double set of brackets, $\la\la\ra\ra$, denotes an average over both particles and events. The flow angles, $\Psi_n$, cannot be measured experimentally event-by-event, so the root-mean-square of the flow coefficients are calculated with two-particle correlations~\cite{Borghini:2001vi}
    \begin{equation}
        \la v_n^2\ra = \la\la \cos n\left(\varphi_1-\varphi_2\right)\ra\ra.
    \end{equation}
The \pt dependence of the flow coefficient is usually studied with the differential flow coefficient $v_n\{2\}(p_\mathrm{T})$~\cite{Borghini:2001vi}
\begin{equation}
v_n\{2\}(p_\mathrm{T}) = \frac{\la\la\cos [n(\varphi_1^{\rm POI}-\varphi_2)]\ra\ra}{\sqrt{\la\la\cos [n(\varphi_1-\varphi_2)]\ra\ra}} = \frac{\la v_n(p_\mathrm{T})~v_n\cos [n(\Psi_n (p_\mathrm{T})-\Psi_n)]\ra}{\sqrt{\la v_n^2\ra}}.
\label{eq:vnpt}
\end{equation}
The $\varphi^{\rm POI}$ and $\varphi$ refer to the azimuthal angles of the particles of interest (POI) and reference flow particles. The $\pt$ refers to the $\pt$ of the POIs selected from narrow $\pt$ ranges. The reference flow particles are chosen from a wide kinematic range, which should ideally be limited to a region dominated by collective behavior. The $\Psi_n(p_\mathrm{T})$ represents the \pt-differential symmetry plane angles at a specific $\pt$ range, which might fluctuate around the reference symmetry plane angles $\Psi_n$. The effect of the difference between $\Psi_n(p_\mathrm{T})$ and $\Psi_n$, due to \pt-dependent flow angle fluctuations, is quantified by the cosine term $\la\cos [n(\Psi_n (p_\mathrm{T})-\Psi_n)]\ra$. The effects of the \pt-dependent flow coefficient fluctuations are observed when the factorization hypothesis is broken
\begin{align}
\la v_n(p_\mathrm{T})v_n\ra \neq \sqrt{\la v_n(p_\mathrm{T})^2\ra}\sqrt{\la v_n^2\ra}.
\end{align}
Aside from the effects of the \pt-dependent fluctuations of the flow angle and the flow magnitude, $\vncr$ also has contributions from non-flow sources such as jets or resonance decays. Such sources provide a flow signal but are not associated with bulk particle production or correlated with the symmetry plane angles $\Psi_n$. 

To account for these effects, another two-particle correlation was proposed in Ref.~\cite{Heinz:2013bua}
\begin{align}
v_n[2](p_\mathrm{T}) &= \sqrt{\la\la\cos [n(\varphi_1^\mathrm{POI}-\varphi_2^\mathrm{POI})]\ra\ra} \\
&= \sqrt{\la v_n(p_\mathrm{T})^2\ra}\nonumber,
\label{eq:vnsq}
\end{align}
that is not affected by fluctuations in the flow angle or flow coefficient, and it is less affected by non-flow effects than \vncr. The difference between \vncr~and \vnsqbr~is that the former takes the reference flow from a wide kinematic range and the POIs from a small $\pt$ interval, and the latter takes two POIs from the same narrow $\pt$ range. Since \vnsqbr~is not affected by the flow angle and flow magnitude fluctuations, the \pt-dependent flow vector fluctuations can be probed by taking the ratio of $v_n\{2\}$ and $v_n[2]$
\begin{equation}
\frac{v_n\{2\}}{v_n[2]} = \frac{\langle v_n(\pt)~v_n\cos n[\Psi_n(\pt)-\Psi_n]\rangle}{\sqrt{\langle v_n(\pt)^2\rangle}\sqrt{\langle v_n^2\rangle}}.
\label{eq:vnratio}
\end{equation}
If the ratio $\vncr/\vnsqbr$ is smaller than unity, it indicates the presence of \pt-dependent flow vector fluctuations. 

Another way to study flow vector fluctuations is to examine the factorization of two-particle correlations from different transverse momentum regions. Factorization of two-particle correlations was observed to hold in some kinematical ranges in Refs.~\cite{Aamodt:2011by, ATLAS:2012at, Chatrchyan:2012wg, PHOBOS:2010ekr}, but is shown not to hold in general in Ref.~\cite{Gardim:2012im}. The factorization can be tested with the factorization ratio $r_n$~\cite{Gardim:2012im}:
\begin{align}
r_n &= \frac{V_{n\Delta}(p_\mathrm{T}^\mathrm{a},p_\mathrm{T}^\mathrm{t})}{\sqrt{V_{n\Delta}(p_\mathrm{T}^\mathrm{a},p_\mathrm{T}^\mathrm{a})V_{n\Delta}(p_\mathrm{T}^\mathrm{t},p_\mathrm{T}^\mathrm{t})}}\nonumber\\
&= \frac{\la v_n(p_\mathrm{T}^\mathrm{a}) v_n(p_\mathrm{T}^\mathrm{t})\cos [n(\Psi_n(p_\mathrm{T}^\mathrm{a})-\Psi_n(p_\mathrm{T}^\mathrm{t})]\ra}{\sqrt{ \la v_n ^2(p_\mathrm{T}^\mathrm{a})\ra\la v_n^2(p_\mathrm{T}^\mathrm{t})\ra}},
\label{eq:rn}
\end{align}
which is a particular case of the ratio shown in Eq. \eqref{eq:vnratio}, obtained by taking particles from two different narrow $\pt$ ranges. Most known sources of non-flow do not factorize at low \pt~\cite{Kikola:2011tu}, so $r_n = 1$ does not always hold. In a system dominated by flow, with no or negligible non-flow effects, $r_n$ is smaller than or equal to unity due to the Cauchy-Schwarz inequality~\cite{Gardim:2012im}. The factorization holds when $r_n$ equals unity, while $r_n$ smaller than unity indicates the presence of \pt-dependent flow vector fluctuations. If the triggered particles are selected from a wide kinematic range (making them equivalent to the reference particles), then $r_n$ becomes identical to $v_n\{2\}/v_n[2]$. In general, however, $r_n$ provides information about the structure of the two-particle correlations for triggered and associated particles, probing the fluctuations of the flow vector at \pta~and \ptt. In contrast, the ratio $v_n\{2\}/v_n[2]$ includes the \pt-integrated information and probes the \pt-differential flow vector with respect to the \pt-integrated flow vector.  

The ratio $v_n\{2\}/v_n[2]$ and the factorization ratio $r_n$ carry information about the flow angle and magnitude fluctuations but cannot isolate both contributions. Thus, it is desirable to separate these two effects to quantify the contributions from each source.\\\\
\noindent
The flow angle fluctuations are studied with the observable \Fpsi{n}, which aims to isolate the \pt-dependent fluctuations of the flow angle~\cite{ALICE:2022dtx}
\begin{align}
\Fpsi{n} = \frac{\la\la\cos [n(\varphi_1^\mathrm{POI}+\varphi_2^\mathrm{POI}-\varphi_3-\varphi_4)]\ra\ra}{\la\la\cos [n(\varphi_1^\mathrm{POI}+\varphi_2-\varphi_3^\mathrm{POI}-\varphi_4)]\ra\ra}&= \frac{\la v_n(\pt)^2~v_n^2 \cos 2n[\Psi_n(\pt)-\Psi_n]\ra}{\la v_n(\pt)^2v_n^2\ra}\nonumber\\
&\simeq \la \cos 2n\left[\Psi_n(\pt)-\Psi_n\right]\ra_w,
\label{eq:refangle}
\end{align}
where the third equality holds if the non-flow contribution is approximately the same for the numerator and denominator. The \textit{w} subscript denotes that $\Fpsi{n}$ is a weighted average with each event having a weight of $v_n^4$~\cite{Bozek:2017qir}. If the flow angle fluctuates as a function of \pt, then $\Fpsi{n}$ will be smaller than unity. If there are no \pt-dependent fluctuations of the flow angle, then $\Fpsi{n}$ is equal to unity. The $\Fpsi{n}$ corresponds to the cosine term in Eq. (\ref{eq:vnratio}) but with twice the angle. Only a lower limit of the single-flow-angle fluctuations, $\la \cos n \left[\Psi_n(\pt)-\Psi_n)\right]\ra$, can be obtained with the trigonometric double-angle formula due to the event averaging
\begin{align}
\sqrt{\frac{\Fpsi{n}+1}{2}} \simeq \sqrt{\la\cos^2n(\Psi_n(\pt)-\Psi_n)\ra} \geq \la\cos n(\Psi_n(\pt)-\Psi_n)\ra.
\label{eq:2AngleAvg}
\end{align}
It is also possible to probe the upper limit on the two-particle flow magnitude fluctuations since it must correspond to the remaining fluctuations of the flow vector. The ratio with $v_n\{2\}/v_n[2]$ quantifies the upper limit of the first-moment flow magnitude fluctuations, since
\begin{align}
\frac{v_n\{2\}/v_n[2]}{\sqrt{\langle \cos^2 n[\Psi_n(p_\mathrm{T}^a)-\Psi_n]\rangle}}
 &\leq \frac{v_n\{2\}/v_n[2]}{\langle \cos n[\Psi_n(p_\mathrm{T}^a)-\Psi_n]\rangle}\nonumber\\ 
 &=\frac{\langle v_n(\pt)~v_n\cos n[\Psi_n(\pt)-\Psi_n]\rangle}{\sqrt{\langle v_n(\pt)^2\rangle}\sqrt{\langle v_n^2\rangle}\langle \cos n[\Psi_n(\pt)-\Psi_n]\rangle\nonumber}\nonumber\\
 &\approx\frac{\la v_n(\pt)~v_n\ra}{\sqrt{\la v_n(\pt)^2\ra}\sqrt{\la v_n^2\ra}}.
\label{eq:Rmag}
\end{align}

The above equation provides an upper limit on the first-moment flow magnitude fluctuations, but an exact measurement of the second-moment flow magnitude fluctuations can be obtained by taking the ratio of the four-particle correlations with opposite signs on the azimuthal angle belonging to particles from the same kinematic region
\begin{align}
\frac{\langle\cos n(\varphi_1^\mathrm{POI}+\varphi_2-\varphi_3^\mathrm{POI}-\varphi_4)\rangle}{\langle\cos n(\varphi_1^\mathrm{POI}-\varphi_3^\mathrm{POI})\rangle\langle\cos n(\varphi_2-\varphi_4)\rangle} = \frac{\langle v_n^2(\pt) v_n^2\rangle}{\langle v_n^2(\pt)\ra\la v_n^2\ra}.
\label{vnmag}
\end{align}
Considering the \pt-integrated $\la v_n^4\ra/\la v_n^2\ra^2$ as the baseline, deviations from such a baseline indicate the presence of \pt-dependent flow magnitude fluctuations. The expression in Eq. \eqref{vnmag} is therefore normalized with the baseline to obtain a double ratio correlator \Fvn{n} for measuring the \pt-dependent flow magnitude fluctuations:
\begin{align}
    \Fvn{n} = \frac{\langle v_n^2(\pt) v_n^2\rangle/\langle v_n^2(\pt)\ra\la v_n^2\ra}{\la v_n^4\ra/\la v_n^2\ra^2}.
\end{align}
Together, these observables, \Fpsi{n} and \Fvn{n}, allow us to probe the flow angle and magnitude fluctuations separately and provide a quantification of both of them. Additionally, the limits extracted with the trigonometric formula are comparable with the previous methods of measuring \pt-dependent flow vector fluctuations~\cite{Acharya:2017ino}. 

\section{Experimental setup and data sample}
\label{sec:exp}
ALICE~\cite{Aamodt:2008zz} is a dedicated heavy-ion experiment at the LHC. One of its focuses is the study of the properties of the QGP. The central barrel of the ALICE detector is encased in a large solenoid magnet. The Inner Tracking System (ITS)~\cite{Aamodt:2010aa} is the innermost detector in the ALICE experiment. Its primary function is to localize the primary vertex with a resolution better than 100 $\mu$m, to reconstruct secondary vertices, and to track and identify low momentum particles (\pt $<$ 200 MeV/$\textit{c}$). It also improves momentum and angle resolution for particles reconstructed by the Time Projection Chamber (TPC)~\cite{Alme:2010ke}. The TPC is the primary tracking detector in ALICE. It is optimized for high-resolution charged-particle momentum measurements ranging from several hundred MeV/\textit{c} up to 100 GeV/\textit{c}. In the TPC, a pseudorapidity coverage of $|\eta|<0.8$ ensures maximal coverage without loss of efficiency at the TPC edges. Additionally, the TPC has full $2\pi$ coverage in the azimuthal direction. The V0 system~\cite{Abbas:2013taa} consists of two arrays, V0A and V0C, which cover the pseudorapidity ranges of $2.8 < \eta < 5.1$ and $-3.7 < \eta < -1.7$, respectively. It is designed to provide triggers for the experiment and to separate beam-beam interactions from the background, such as beam-gas interactions. It is also used to measure charged-particle multiplicity in the forward region, which is used to determine the centrality of nucleus-nucleus collisions~\cite{ALICE:2013hur}. 

The events are selected according to a minimum bias trigger criterion which requires at least two of the following~\cite{Abbas:2013taa}: 1) hits in the Silicon Pixel Detector of the ITS, 2) a signal in V0A, and 3) a signal in the V0C, as well as a reconstructed primary vertex within $\pm10$ cm of the nominal interaction point along the beam axis. The centrality of the events is determined by the sum of V0 signal amplitude in the scintillator arrays V0A and V0C~\cite{ALICE:2013hur}. Pileup events refer to events that are contaminated by one or more out-of-bunch or in-bunch collisions occurring within the readout time of the TPC. Such contaminated events cannot be accurately assigned to a proper centrality interval. Pileup events are therefore rejected based either on the presence of multiple reconstructed vertices or on the correlations between the number of tracks measured in the TPC and the number of tracks reconstructed with relatively fast detectors such as the ITS and Time-of-Flight (TOF)~\cite{Abelev:2014ffa}. This effect is most significant in central collisions due to the large multiplicity of the pileup events. In this paper, 54M \PbPb collisions at $\snnPb$ measured in the 2015 data-taking period at the LHC pass the event selection criteria. Charged tracks are reconstructed using the ITS and the TPC. Tracks are selected with at least 70 TPC space points out of a maximum of 159 possible points and a $\chi^2$ per degree of freedom of the track fit to TPC space points less than $4$. Tracks are required to have at least one hit in the Silixon Pixel Detector (SPD). Additionally, tracks must have a distance of closest approach (DCA) to the primary vertex of less than $2~\mathrm{cm}$ in the longitudinal direction and a \pt-dependent selection in the transverse direction ranging from $0.2~\mathrm{cm}$ at $0.2~\mathrm{GeV/\textit{c}}$ to $0.016~\mathrm{cm}$ at $5~\mathrm{GeV/\textit{c}}$. Finally, the charged tracks are taken from the kinematic range of $|\eta| <0.8$, and the tracks used for reference particles are also within $0.2<\pt^\mathrm{ref}<5.0~\mathrm{GeV/\textit{c}}$. The $\pt$ range is selected to extend beyond the upper bound of validity for hydrodynamics ($\sim 3~\mathrm{GeV/\textit{c}}$) as the flow angle and magnitude fluctuations may increase beyond this regime. Non-flow correlations are suppressed by requiring a pseudorapidity gap, $|\Delta\eta|$, greater than or equal to 0.8 between particles in the calculation of the flow coefficients with two-particle correlations, and a subevent method with no pseudorapidity gap is used in the calculation of four-particle correlations.

\section{Statistical and systematic uncertainties}
\label{sec:unc}
The statistical uncertainties of the measurements are estimated with the bootstrap method of random sampling with replacement~\cite{EfronBS}. Ten similarly sized subsamples are sampled uniformly from the entire event ensemble. From these ten subsamples, 1000 generated event samples are constructed by randomly selecting ten subsamples from the original ten subsamples with replacement, i.e., the same subsample can be selected multiple times. For each of the 1000 generated event ensembles, the observables are calculated as a weighted average, providing a distribution for each observable. The statistical uncertainty is then estimated from the variance of the distribution for a given observable, which should approach the actual distribution given a large enough sampling.
\begin{table}[h!]
\caption{Systematic uncertainties estimated from variations of event and track selection criteria. The uncertainties may vary with centrality and are, in those cases, given as a lower and upper bound. Systematic uncertainties that are not statistically significant are listed as N/S. See text for details.}
\begin{center}
\begin{tabular}{c|c|c|c|c|c|c|c}
    &$v_2\{2\}/v_2[2]$  &$v_3\{2\}/v_3[2]$  &$v_4\{2\}/v_4[2]$  &$r_2$  &$r_3$  &\Fpsi{2}   &\Fvn{2}\\\hline
Cent. est. &N/S   &0-0.2\%   &0-0.7\%   &0-0.3\%   &0-2.4\%    &0-0.1\%   &0-0.1\%\\\hline
$|V_z|$ &N/S   &N/S   &N/S   &0-0.1\%   &0-0.6\%   &0-0.1\%   &0-0.1\%\\\hline
Mag. field &0-0.1\%   &0.1-1\%   &0-2.4\%   &0-0.5\%   &0-2\%    &0.4\%   &0.4\%\\\hline
Pileup &N/S   &0-0.3\%   &0-1.2\%   &0-0.5\%   &0-1.1\%    &N/S   &N/S\\\hline
\# TPC cls. &N/S   &N/S   &0-0.7\%   &0-2\%   &0-2\%   &0.7\%   &0.7\%\\\hline
Track type &0-0.2\%   &0.5-1.3\%   &1.1\%-2.2\%   &0-1.8\%   &0-1.3\%    &0-0.1\%   &0-0.1\%\\\hline
$|\mathrm{DCA}_{z}|$ &N/S   &N/S   &N/S   &N/S   &N/S    &N/S   &N/S\\\hline
$|\mathrm{DCA}_{xy}|$  &N/S   &0-1\%   &N/S   &N/S   &0-1\% &0-0.1\%   &0-0.1\%\\\hline
$\chi^2$ TPC cls.   &0-0.1\%   &0-0.3\%   &0-3.1\%   &0-0.8\%   &0-0.5\%  &0.4\%   &0.4\%\\\hline
Non-flow    &0-0.6\%   &0.6\%-1.7\%   &2\%-3.4\%   &0.1\%-2.3\%   &0-5.9\%   &1.1\%   &1.1\%\\\hline
\end{tabular}
\end{center}
\label{table:sys}
\end{table}

The systematic uncertainties of the measurements are evaluated by varying the event and track selection criteria and are shown in Table~\ref{table:sys}. The systematic uncertainties related to the event selection are investigated by repeating the analysis using different detectors for the centrality determination, changing the selection on the position of the primary vertex along the beam direction $|V_{z}|$, testing different magnetic field polarities, and testing different pileup selections. The systematic uncertainty associated with the centrality determination (Cent. est.) is estimated by conducting the full analysis with the SPD as an alternative centrality estimator. It is negligible for most observables but contributes up to 2.4\% for $r_3$. The systematic uncertainty related to different primary vertex position criteria ($|V_z|$) is studied by changing the criterion from $|V_{z}| < $ 10 cm to $|V_{z}| < $ 7 cm, 8 cm, and 9 cm and is found to contribute at most up to 0.6\% for $r_3$. The effects of the magnetic field polarity (Mag. Field) are tested by analyzing datasets with different magnetic configurations and yield a systematic uncertainty ranging from negligible and up to 2.4$\%$ for $v_4\{2\}/v_4[2]$. The systematic effect of pileup is estimated by changing the pileup selection in centrality interval 0-10\% (Pileup), where the pileup events are expected to have the largest impact due to large multiplicities in the TPC. The pileup selection variation contribution to the systematic uncertainty ranges from negligible to at most $\sim$1\% for $r_3$ and $v_4\{2\}/v_4[2]$. 

The quality of the reconstructed tracks is varied by changing the track type to include tracks without hits in the SPD (Track type) and by modifying the minimum number of TPC space points required (\# TPC cls.) from $70$ to $80$ and $90$. This variation yields up to a 2\% systematic uncertainty in the $r_n$ observables and less than 1\% in the other observables. Additionally, the requirement of maximum DCA in the longitudinal ($z$) direction ($|\mathrm{DCA}_z|$) is changed from 2 cm to 0.5 and 1 cm, and in the transverse ($xy$) direction ($|\mathrm{DCA}_{xy}|$) it is changed from a \pt-dependent selection ($|\mathrm{DCA}_\mathrm{xy}| \leq N_\sigma \times (0.0015+0.005/\pt^{1.1})$) corresponding to $7\sigma$ deviation from the expected functional form to one corresponding to $4\sigma$. Both variations of the $|\mathrm{DCA}_z|$ and the variation of $|\mathrm{DCA}_{xy}|$ selection criteria yield negligible contributions to the systematic uncertainty for most of the observables. The last variation considered for the track quality is the $\chi^2$ per TPC cluster ($\chi^2$ TPC cls.), which is tightened from 4 to 2.5 and yields a systematic uncertainty of up to 3.1\% for $v_4\{2\}/v_4[2]$ but less than half a percent for the other observables. To estimate the systematic uncertainty associated with the non-flow suppression in two-particle correlations (Non-flow), the analysis is repeated with pseudorapidity gaps of $|\Delta\eta| < 0.6, 1.0, 1.2$. The consistency between results with different pseudorapidity gaps suggests that short-range non-flow correlations are suppressed. It is possible that remaining long-range non-flow correlations such as those from momentum conservation and di-jets could influence the results, even though an additional Monte Carlo study with HIJING~\cite{Wang:1991hta}, a heavy-ion model that does not contain collective effects, showed results for the correlators in Eqs. \eqref{eq:refangle} and \eqref{eq:Rmag} consistent with zero. Based on the above studies, it is found that the remaining non-flow contribution to the observables is less than $\sim 2\%$ for the elliptic flow observables. The systematic uncertainty is estimated for each centrality interval separately (and additionally for each $\ptt$ range for the calculation of $r_n$). The statistical significance of the systematic uncertainty is evaluated with the Barlow check introduced in Ref.~\cite{Barlow:2002yb}. Only systematic uncertainties found to be statistically significant according to this check are considered for the final systematic uncertainty. The total systematic uncertainty is calculated as the quadratic sum of the individual sources. Only the variation resulting in the largest uncertainty is added to the total systematic uncertainty for sources with more than one variation.

\begin{figure}[htbp]
\centering
\includegraphics[scale=0.8]{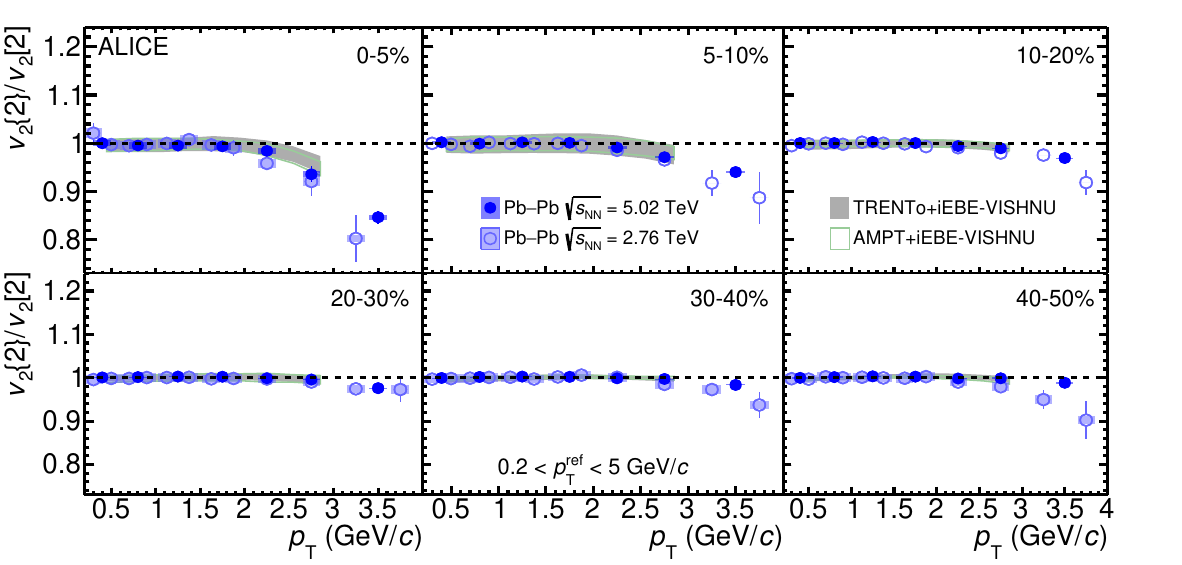}
\caption{The ratio $v_2\{2\}/v_2[2]$ in $\PbPb$ collisions at $\snn$ = 5.02 TeV (solid dark blue circles) and 2.76 TeV~\cite{Acharya:2017ino} (open light blue circles) as a function of transverse momentum. The different panels display results in different centrality intervals. Statistical (systematic) uncertainties are represented by solid bars (faded boxes). Predictions from the iEBE-VISHNU hydrodynamic model with T$_{\rm R}$ENTo initial conditions and temperature-dependent $\eta/s(T)$~\cite{Zhao:2017yhj}, and with AMPT initial conditions and $\eta/s = 0.08$~\cite{Zhao:2017yhj}, are shown in colored bands.}
\label{res:v2ratio}
\end{figure}
\section{Results}
\label{sec:res}
In this paper, precision measurements of the ratio $v_n\{2\}/v_n[2]$ are presented for $n = 2,3$ and $4$ in $\PbPb$ collisions at $\snnPb$. The results are compared with the existing measurements at $\snnPbOne$. Figure~\ref{res:v2ratio} shows the ratio $v_2\{2\}/v_2[2]$ with $|\Delta\eta| > 0.8$ as a function of $\pt$ for various centrality intervals ranging from 0--5\% up to 40--50\%. For the most central collisions (0--5\%), the ratio for $n=2$ is consistent with unity up to $\pt\approx 2$ GeV/\textit{c}. It starts to deviate from unity as the $\pt$ increases with a significance higher than $2\sigma$, $3\sigma$, and $5\sigma$ in the three bins above 2 GeV/\textit{c}, respectively. The ratio reaches a deviation of 15\% from unity at $\pt > 3~\mathrm{GeV/\textit{c}}$. For centrality intervals larger than 20\%, the ratios are close to unity within 2\% for the presented \pt range. This trend was already observed with measurements based on ALICE data at \snn = 2.76 GeV/\textit{c}~~\cite{Acharya:2017ino}.  
The data are compared to several theoretical models. The iEBE-VISHNU model is a (2+1)D event-by-event relativistic viscous hydrodynamic model coupled to a hadronic cascade model~\cite{Shen:2014vra}. In this paper, two sets of calculations with the iEBE-VISHNU model are used: one with \trento~\cite{Moreland:2014oya} and one with AMPT~\cite{Lin:2004en} initial conditions. The model calculations with \trento initial conditions use a temperature-dependent specific shear viscosity $\eta/s(T)$, while the calculations with AMPT initial conditions use a $\eta/s=0.08$. The input parameters of iEBE-VISHNU are tuned according to~\cite{Zhao:2017yhj}. The hydrodynamic calculations are performed in a $p_\mathrm{T}$ range up to 3 GeV/\textit{c}, as this is the region where hard processes are expected to take over. 

The theory curves describe quantitatively  $v_2\{2\}/v_2[2]$ within the uncertainties for both sets of model calculations. The large uncertainties of the hydrodynamic calculations are due to limited number of produced Monte Carlo events. 

\begin{figure}[htbp]
\centering
\includegraphics[scale=0.8]{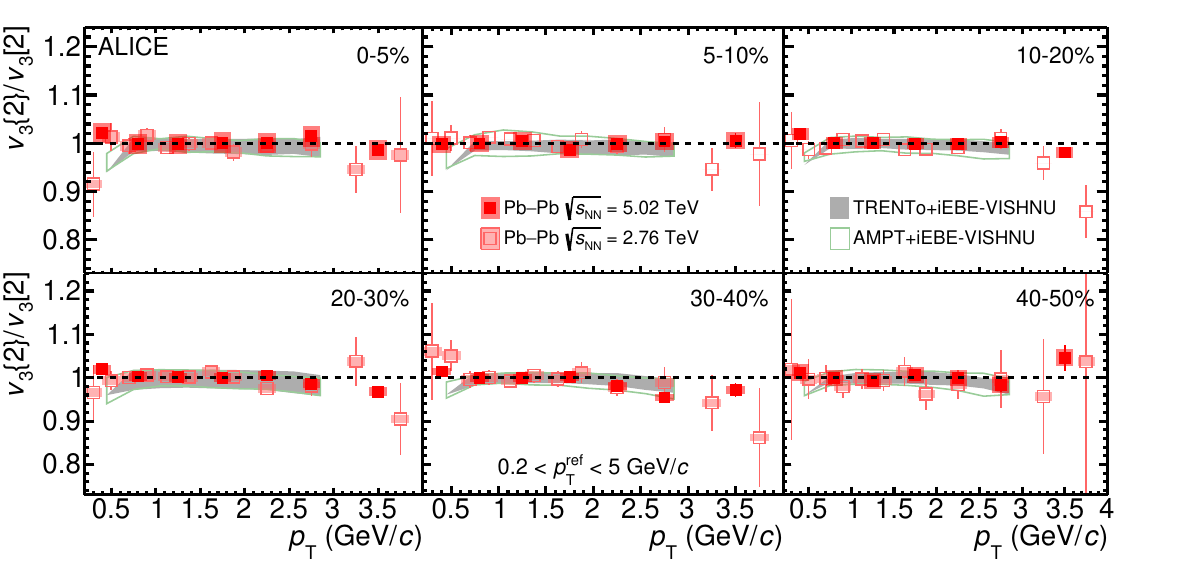}
\caption{The ratio $v_3\{2\}/v_3[2]$ for $\PbPb$ collisions at $\snn$ = 5.02 TeV (solid dark red squares) and 2.76 TeV~\cite{Acharya:2017ino} (open light red squares) as a function of transverse momentum. The different panels display results in different centrality intervals. Statistical (systematic) uncertainties are represented by solid bars (faded boxes). Predictions from iEBE-VISHNU hydrodynamic model with T$_{\rm R}$ENTo initial conditions and temperature-dependent $\eta/s(T)$~\cite{Zhao:2017yhj}, and with AMPT initial conditions and $\eta/s = 0.08$~\cite{Zhao:2017yhj}, are shown in colored bands.}
\label{res:v3ratio}
\end{figure}
\begin{figure}[htbp]
\centering
\includegraphics[scale=0.8]{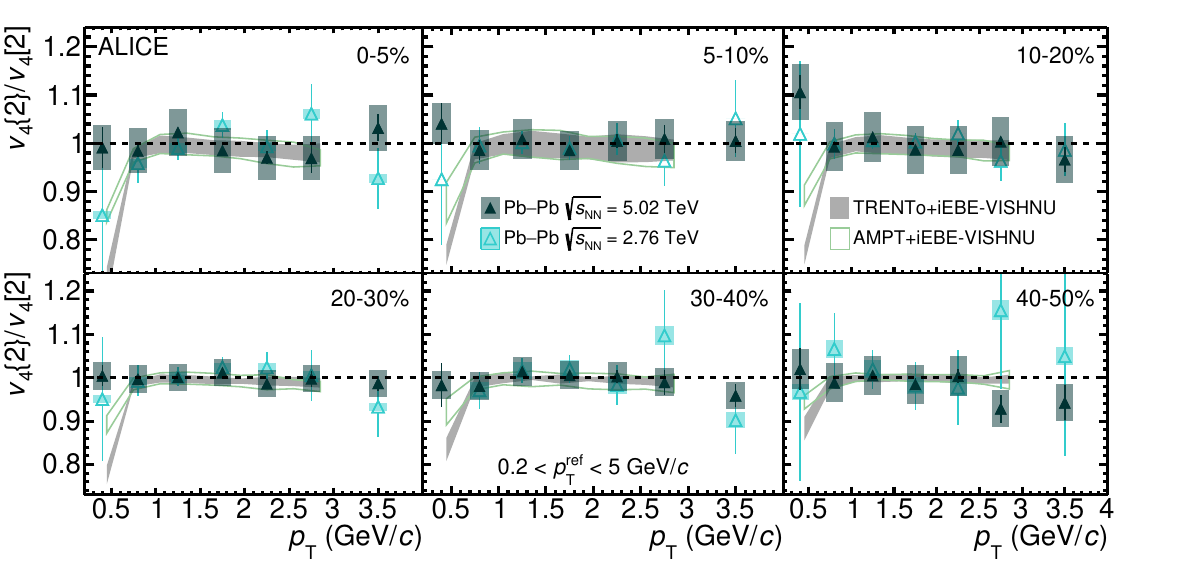}
\caption{The ratio $v_4\{2\}/v_4[2]$ for $\PbPb$ collisions at $\snn$ = 5.02 TeV (solid dark cyan triangles) and 2.76 TeV~\cite{Acharya:2017ino} (open light cyan triangles) as a function of transverse momentum. The different panels display results in different centrality intervals. Statistical (systematic) uncertainties are represented by solid bars (faded boxes). Comparison with iEBE-VISHNU hydrodynamic model with T$_{\rm R}$ENTo initial conditions and temperature-dependent $\eta/s(T)$~\cite{Zhao:2017yhj}, and with AMPT initial conditions and $\eta/s = 0.08$~\cite{Zhao:2017yhj}, are shown in colored bands.}
\label{res:v4ratio}
\end{figure}

Higher-order anisotropic flow measurements were measured for the first time in Ref.~\cite{ALICE:2011ab} and were found to be more sensitive to the initial conditions and properties of the QGP~\cite{Alver:2010dn}. The ratio $v_3\{2\}/v_3[2]$ with $|\Delta\eta| > 0.8$ is shown in Fig.~\ref{res:v3ratio}. It can be seen that the ratio agrees with unity in the presented centrality and \pt~ranges, unlike  $v_2\{2\}/v_2[2]$, as shown in Fig.~\ref{res:v2ratio}. The agreement with unity suggests that the triangular flow vector $\Vn{3}$ does not fluctuate strongly with \pt~in the presented \pt~and centrality ranges. Previously published measurements~\cite{Acharya:2017ino} have substantial uncertainties for $v_3\{2\}/v_3[2]$ and found no significant $\Vn{3}$ fluctuations. With these results, the findings in Ref.~\cite{Acharya:2017ino} are confirmed with substantially increased statistics, and it can be concluded that there are no significant \pt-dependent $\Vn{3}$ fluctuations in $\PbPb$ collisions at $\snnPb$ within the current experimental uncertainties. The hydrodynamic calculations with the iEBE-VISHNU hydrodynamic models describe the data. The models with T$_{\rm R}$ENTo and AMPT initial conditions show agreement with unity and the data. At small \pt in central collisions, the hydrodynamical calculations deviate slightly from unity and overestimate the effect of the flow vector fluctuations observed in the data. 

\begin{figure}[htbp]
\centering
\includegraphics[scale=0.8]{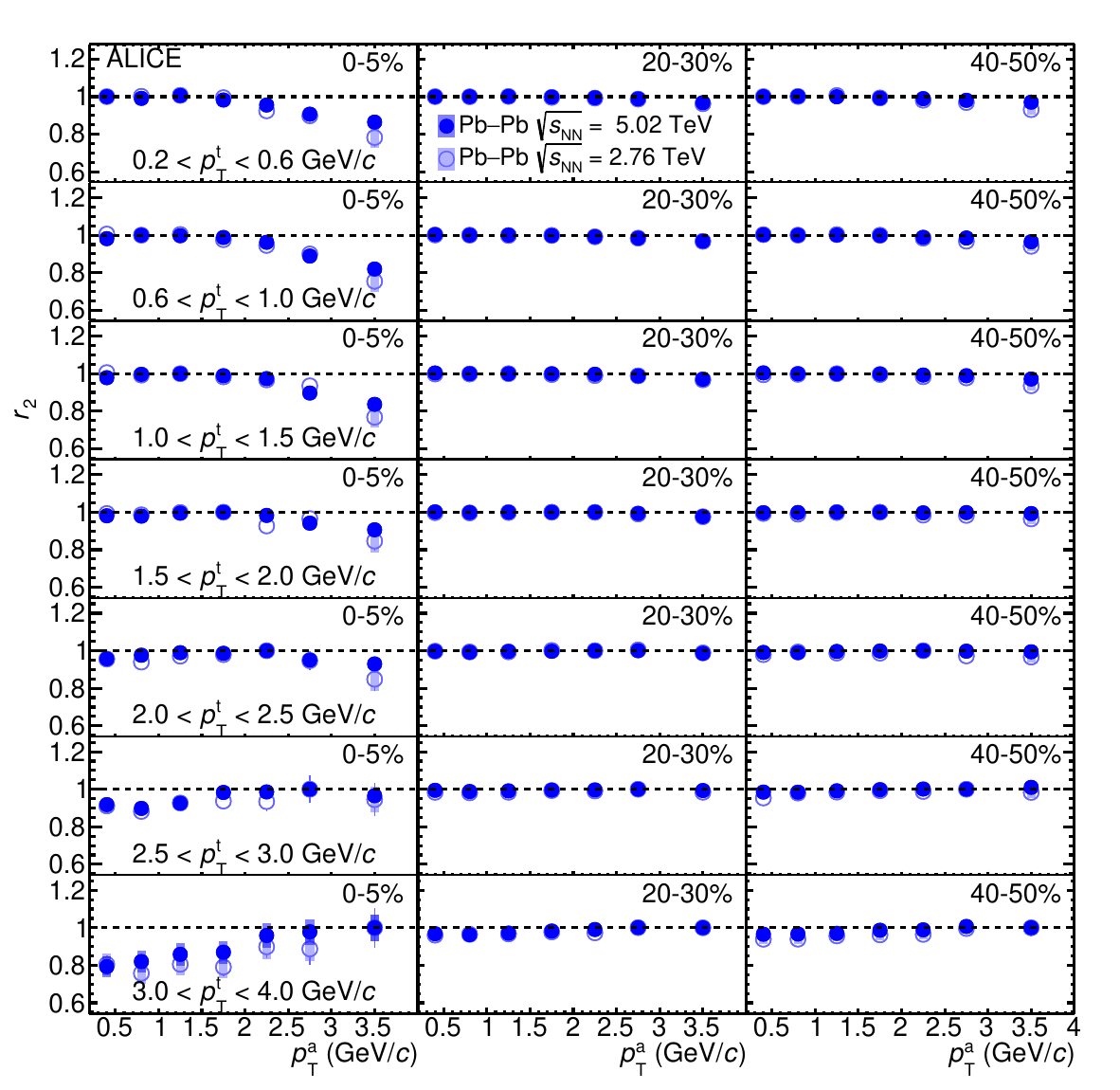}
\caption{The factorization ratio $r_2$ for $\PbPb$ collisions at $\snn$ = 5.02 TeV (dark blue circles) and 2.76 TeV~\cite{Acharya:2017ino} (light blue circles) as a function of associated particle $\pta$. The columns show the results in centrality intervals 0--5\%, 20--30\%, and 40--50\%, while the rows show the results for different trigger particle $p_\mathrm{T}^t$ intervals. Statistical (systematic) uncertainties are represented by solid bars (faded boxes).}
\label{res:r2Pt}
\end{figure}

The ratio $v_4\{2\}/v_4[2]$ with $|\Delta\eta| > 0.8$ shown in Fig.~\ref{res:v4ratio} is consistent with unity within the uncertainties across all centrality intervals. The previous measurements of $v_4\{2\}/v_4[2]$~\cite{Acharya:2017ino} had large statistical uncertainties and showed no statistically significant deviation from unity. The results presented in this paper do not show any sign of significant fluctuations of $\Vn{4}$ as a function of transverse momentum with significantly smaller uncertainties compared to the measurements in Ref.~\cite{Acharya:2017ino}. The hydrodynamic calculations are consistent with unity for $p_\mathrm{T} > 0.6$ GeV/\textit{c} but show a deviation from unity at low \pt inconsistent with the measured $v_4\{2\}/v_4[2]$.

This paper also presents precision measurements of the factorization ratios $r_n$ in \PbPb collisions at $\snnPb$ for $n=2$ and $3$, calculated according to Eq. \eqref{eq:rn}. Figure~\ref{res:r2Pt} shows $r_2$ with a pseudorapidity gap $|\Delta\eta|>0.8$ as a function of \pta in centrality intervals 0--5\%, 10--20\%, and 40--50\% in various bins of \ptt. For all $\ptt$ bins, it is observed that the deviations from unity are largest in central collisions, where the initial-state geometry fluctuations dominate, and that the effect becomes more pronounced as the difference $|p_\mathrm{T}^a-p_\mathrm{T}^t|$ increases. The largest deviations from unity are observed in central collisions for $0.2< \ptt < 0.6~\mathrm{GeV/\textit{c}}$ with $3.0<\pta<4.0~\mathrm{GeV/\textit{c}}$ (first row, left panel) and for $3.0<\ptt<4.0~\mathrm{GeV/\textit{c}}$ with $0.2<\pta<0.6~\mathrm{GeV/\textit{c}}$ (last row, left panel), since this is the momentum region where the difference $|\pta-\ptt|$ is the largest. For 40--50\% centrality, the deviation from unity is at most 3\% across the different $\ptt$ ranges. The factorization is broken in central collisions, which, in turn, implies the presence of \pt-dependent flow vector fluctuations as described in Ref.~\cite{Gardim:2012im}. At a higher $\ptt$, the deviations from unity become less pronounced since the difference $|\pta-\ptt|$ reaches the largest value in the lowest $\ptt$ bin. Significant deviations of $r_2$ from unity have been measured at lower energy~\cite{Acharya:2017ino} and confirmed here in $\PbPb$ collisions at $\snn$ = 5.02 TeV. Compared with previous results, the precision of $r_2$ is drastically improved, with the deviations from unity being significant to more than 5$\sigma$ at $3.0 < \pta < 4.0$ GeV/\textit{c} across the presented centralities.

\begin{figure}[htbp]
\centering
\includegraphics[scale=0.8]{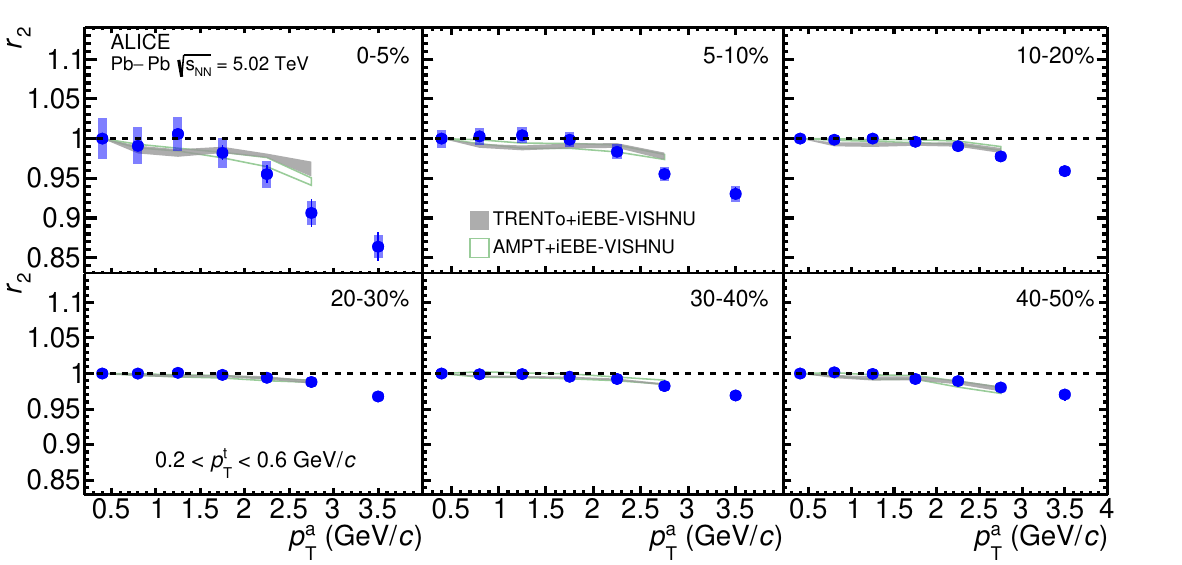}
\caption{The factorization ratio $r_2$ for $\PbPb$ collisions at $\snn$ = 5.02 TeV (blue circles) as a function of $\pta$ for $0.2 < \ptt < 0.6$ GeV/\textit{c}. The different panels display results in different centrality intervals. Statistical (systematic) uncertainties are represented by solid bars (faded boxes). Predictions from iEBE-VISHNU hydrodynamic model with T$_{\rm R}$ENTo initial conditions and temperature-dependent $\eta/s(T)$~\cite{Zhao:2017yhj}, and with AMPT initial conditions and $\eta/s = 0.08$~\cite{Zhao:2017yhj}, are shown in colored bands.}
\label{res:r2}
\end{figure}

The centrality dependence of $r_2$ is more clearly seen in Fig.~\ref{res:r2}, where $r_2$ is presented in the centrality intervals 0--5\% to 40--50\% in the lowest $\ptt$ bin of $0.2 < \ptt < 0.6$ GeV/\textit{c}. The comparison with the hydrodynamic calculations from iEBE-VISHNU with AMPT and T$_{\rm R}$ENTo initial conditions is presented. Both hydrodynamic calculations qualitatively describe the trend of $r_2$. However, they also underestimate the deviations from unity at higher $\pt$ in central collisions. The hydrodynamic model with AMPT initial conditions produces a slightly larger deviation of $r_2$ from unity at $\pt>2.5~\mathrm{GeV/\textit{c}}$ in central collisions than the one with T$_{\rm R}$ENTo initial conditions, while both provide a reasonable description of the data in peripheral collisions.

\begin{figure}[t]
\centering
\includegraphics[scale=0.8]{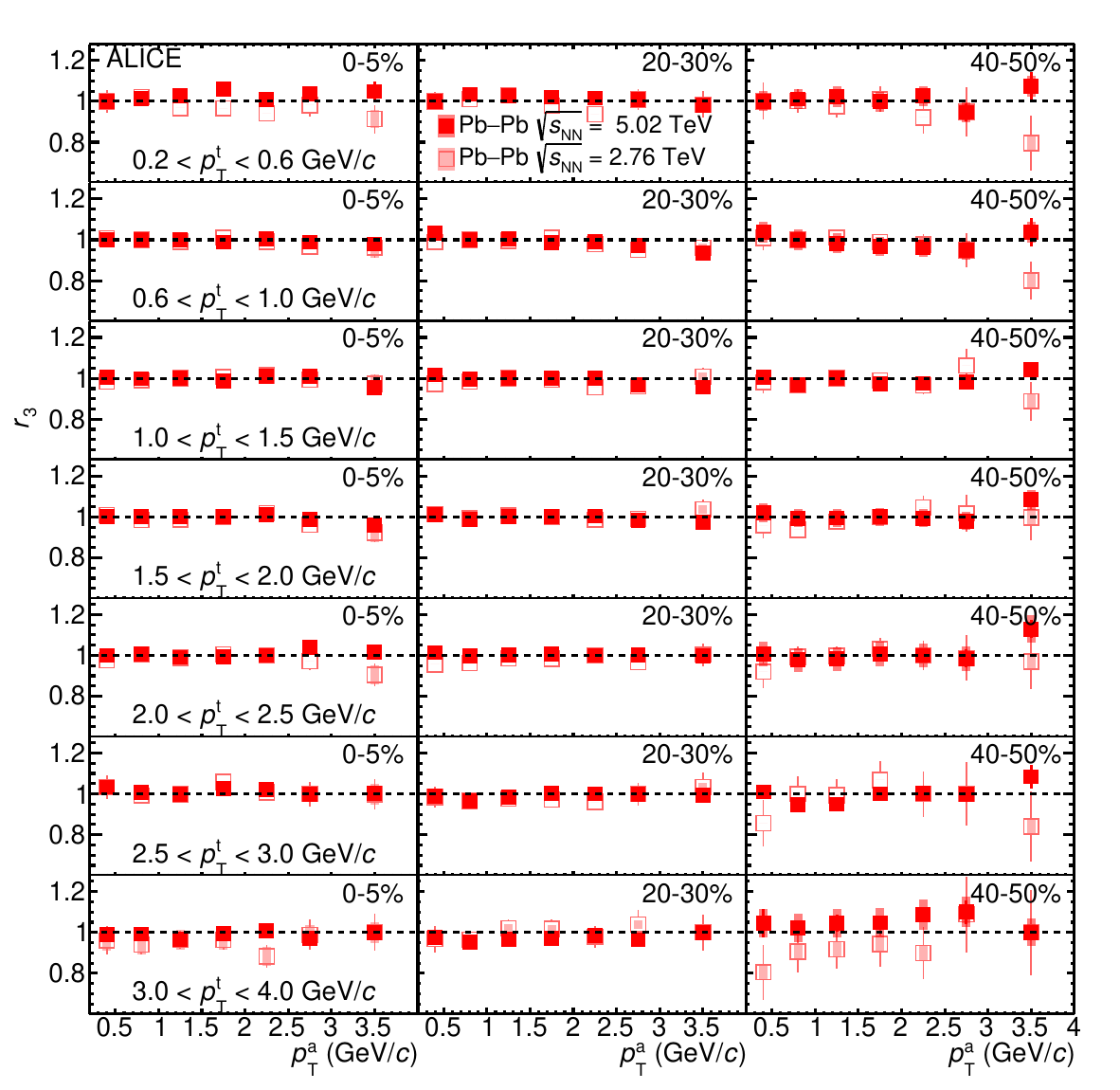}
\caption{The factorization ratio $r_3$ in $\PbPb$ collisions at $\snn$ = 5.02 TeV (solid dark red squares) and 2.76 TeV~\cite{Acharya:2017ino} (open light red squares) as a function of $\pta$. The columns show the results in centrality intervals 0--5\%, 20--30\%, and 40--50\%, while the rows show the results for different trigger particle $p_\mathrm{T}^t$ intervals. Statistical (systematic) uncertainties are represented by solid bars (faded boxes).}
\label{res:r3Pt}
\end{figure}

Figure~\ref{res:r3Pt} shows $r_3$ with $|\Delta\eta| > 0.8$ as a function of $\pta$ for different bins of \ptt, and in centrality intervals 0--5\%, 10--20\%, and 40--50\%. Here, $r_3$ is consistent with unity in the presented centralities and $\pta$ range for all \ptt. The agreement with unity over the presented centrality range suggests no significant $\Vn{3}$ fluctuations independently of the centrality. The lack of a centrality dependence agrees with the picture that triangular flow is driven by initial-state fluctuations rather than the average geometry. The factorization is also observed to hold over the presented ranges of $\pta$ and $\ptt$, as opposed to $r_2$. The previous measurements~\cite{Acharya:2017ino} showed deviations from unity at high \pt~in several bins of \ptt but without a large significance (less than $3\sigma$). It was noted that a possible breakdown of the factorization would be within 10\% when both $\pta$ and $\ptt$ are below 3 GeV/\textit{c}. The precision measurements presented in this paper lowers the possible breakdown of factorization of the triangular flow, $v_3$, down to 1\% across the presented \pt and centrality ranges within a 95\% confidence interval.

\begin{figure}[htbp]
\centering
\includegraphics[scale=0.8]{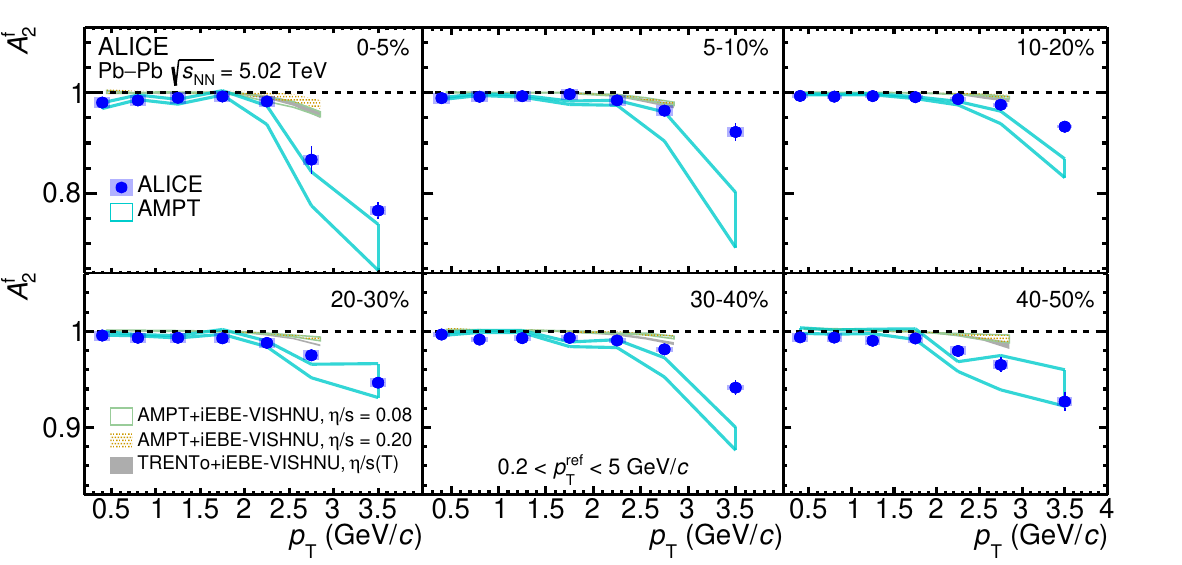}
\caption{The flow angle fluctuations $\Fpsi{2}$ in $\PbPb$ collisions at $\snn$ = 5.02 TeV (blue circles) as a function of $\pt$. The different panels display results in different centrality intervals. Statistical (systematic) uncertainties are represented by solid bars (faded boxes). Predictions from iEBE-VISHNU with AMPT initial conditions and $\eta/s = 0.08, 0.20$ and iEBE-VISHNU with \trento initial conditions and $\eta/s(T)$~\cite{Zhao:2017yhj} as well as AMPT~\cite{Lin:2004en} are shown in colored bands. }
\label{res:FPsi2}
\end{figure}	

The measurements of the \pt-dependent flow angle fluctuations \Fpsi{2} are shown in Fig.~\ref{res:FPsi2} as a function of the transverse momentum $\pt$ in centrality intervals 0--5\% to 40--50\%. More than a $5\sigma$ significance is observed in all centralities for the flow angle fluctuations at the highest $\pt$ values. A deviation from unity of up to 23\% is observed in the 5\% most central collisions for the highest $\pt$ value with a significance of $13\sigma$. The strength of the fluctuations decreases towards more peripheral collisions to around 5\% in 20--30\% and 30--40\% and then slightly increases in 40--50\% centrality up to 7\%. These measurements provide evidence of \pt-dependent flow angle fluctuations. As the systematic uncertainty accounts for any potential remaining non-flow contributions, it is clear that they cannot explain the deviation of \Fpsi{2} from unity. Large fluctuations in central collisions are expected since the collisions are dominated by event-by-event fluctuations in the position of the nucleons and of the quarks and gluons within the nucleons. For the more peripheral events, the pressure gradients due to the geometric anisotropy of the overlap region between the colliding nuclei dominate, decreasing the flow angle fluctuations. The comparison with hydrodynamic calculations shows that iEBE-VISHNU model underestimates the effects of the flow angle fluctuations at $\pt>2.5~\mathrm{GeV/\textit{c}}$. The \trento+iEBE-VISHNU with temperature-dependent shear viscosity to entropy density ratio $\eta/s(T)$ and AMPT+iEBE-VISHNU with $\eta/s=0.08$ show around 5\% (4\%) deviation from unity with a significance of 7$\sigma$ (9$\sigma$) in the 0-5\% central collisions compared to AMPT+iEBE-VISHNU with $\eta/s = 0.20$, which predicts around 2\% deviation from unity with 2.7$\sigma$ significance. However, none of the hydrodynamic models succeeds in quantitatively describing the large measured flow angle fluctuations at high $\pt$ in the \cent{0}{5} central collisions. The AMPT transport model calculations quantitatively predict the flow angle fluctuations in the \cent{0}{5} most central collisions and the \cent{20}{30} and \cent{40}{50} centrality intervals but overestimate the deviation in the other centrality intervals.  

\begin{figure}[htbp]
\centering
\includegraphics[scale=0.8]{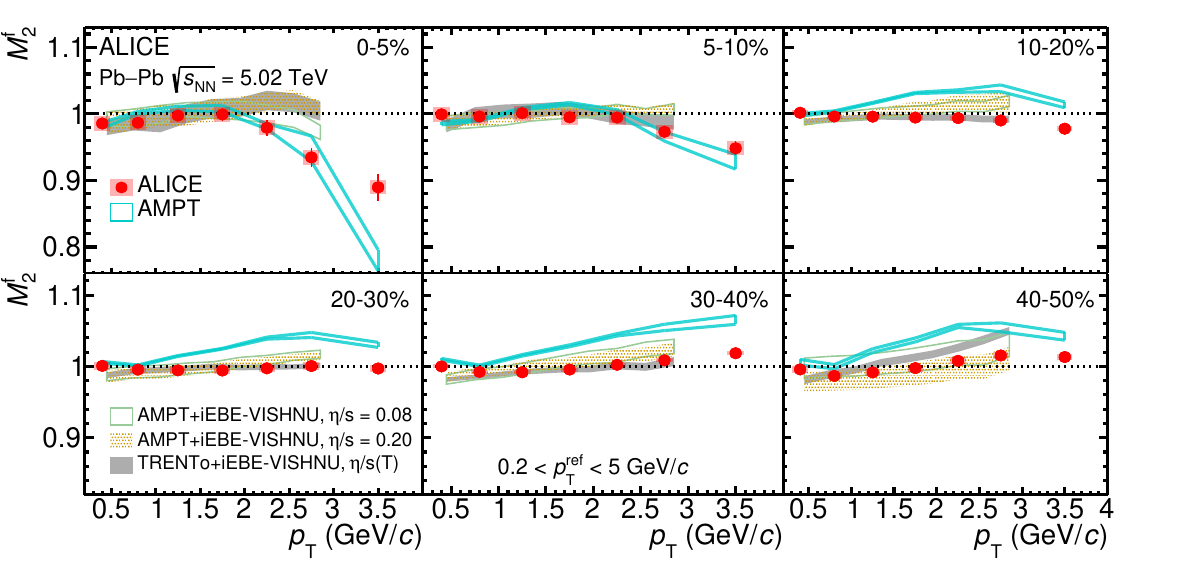}
\caption{The flow magnitude fluctuations $\Fvn{2}$ in $\PbPb$ collisions at $\snn$ = 5.02 TeV (red squares) as a function of $\pt$. The different panels display results in different centrality intervals. Statistical (systematic) uncertainties are represented by solid bars (faded boxes). Predictions from iEBE-VISHNU with AMPT initial conditions and $\eta/s = 0.08, 0.20$ and iEBE-VISHNU with \trento initial conditions and $\eta/s(T)$~\cite{Zhao:2017yhj} as well as AMPT~\cite{Lin:2004en} are shown in colored bands.  }
\label{res:Fv2}
\end{figure}
\begin{figure}[htbp]
\centering
\includegraphics[scale=0.8]{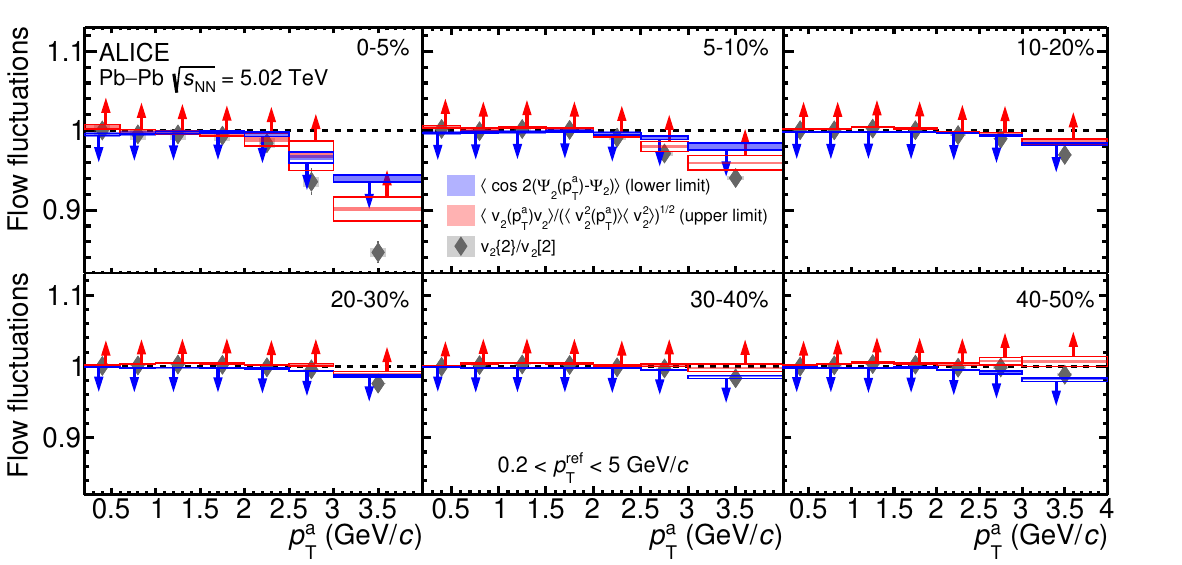}
\caption{The lower limit of $\la\cos n(\Psi_2(\pt)-\Psi_2)\ra$ (blue boxes), the upper limit of $\la v_2(\pt)v_2\ra/\sqrt{\la v_2^2(\pt)\ra\la v_2^2\ra}$ (red boxes), and the flow vector fluctuations $v_{2}\{2\}/v_2[2]$ (gray diamonds) as a function of $\pt$. The different panels display results in different centrality intervals. The red (blue) arrows denote the upper (lower) limits, and the statistical (systematic) uncertainties are represented by open (shaded) boxes.}
\label{res:FlowLims}
\end{figure}

Figure~\ref{res:Fv2} shows the measurements of \Fvn{2} as a function of the transverse momentum $\pt$ in centrality intervals 0--5\% to 40--50\%. Similarly to the flow angle fluctuations, a significant deviation from unity is observed in the 0--5\% most central collisions with values up to $\sim12\%$. As the centrality increases to $\cent{20}{30}$, the flow magnitude fluctuations become smaller. Towards more peripheral collisions, \Fvn{2} becomes larger than unity. In two-particle correlations, the values of ratios such as $\vncr/\vnsqbr$ and $r_n$ are strictly less than unity due to the Cauchy-Schwarz inequality. However, by construction, \Fvn{2} does not satisfy this inequality, and so values can exceed unity. The comparison with theoretical predictions shows that iEBE-VISHNU with \trento initial conditions and AMPT initial conditions underestimate the deviations in the most central collisions. At higher centralities the model calculations are consistent with the data and correctly describe the modestly increasing trend in the data in 30--40\% and 40--50\% centrality intervals. The comparison of the AMPT+iEBE-VISHNU hydrodynamic calculations with different $\eta/s$ show some difference in the 5\% most central collisions. Such a dependence of \Fvn{2} on $\eta/s$ in hydrodynamic models was also seen in Ref.~\cite{ALICE:2022dtx}, while further AMPT model calculations~\cite{Nielsen:2022jms} suggest that \Fvn{2} is not very sensitive to the value of $\eta/s$. This should be investigated with further model studies. Pure AMPT calculations accurately describe the deviation from unity in the \cent{0}{5} and \cent{5}{10} most central collisions but fail to reproduce the data at higher centralities. 

In Fig.~\ref{res:FlowLims}, the lower limit of the flow angle fluctuations (see Eq.~\ref{eq:2AngleAvg}), the upper limit of the flow magnitude fluctuations (see Eq.~\ref{eq:Rmag}) and the flow vector fluctuations (see Eq.~\ref{eq:vnratio}) are shown as a function of transverse momentum in centrality intervals 0--5\% to 40--50\%. The figure shows the limits of the single flow angle and first-moment flow magnitude fluctuations, which are the factors of the total flow vector fluctuations measured by \vncr/\vnsqbr. While the exact contribution of the single-angle flow angle fluctuations and the first-moment flow magnitude fluctuations cannot be established, information about the fluctuations can still be inferred from the limits. The flow angle fluctuations contribute at least $\sim$40\% to the total flow vector fluctuations measured with $v_2\{2\}/v_2[2]$ in 0--5\% central collisions, and it also contributes at higher centralities, where all the sources of fluctuations decrease. This effect is consistent with the measured $\Fpsi{2}$ and with what has been predicted by hydrodynamic calculations~\cite{Heinz:2013bua}. At centralities above 30\%, the first moments of the flow magnitude fluctuations vanish, as the upper limit of the flow magnitude fluctuations, denoted by the red squares in Fig.~\ref{res:FlowLims}, converges towards the lower limit of unity (with unity implying no fluctuations). The disappearance of the flow magnitude fluctuations indicates that above 30\% centrality, the flow vector fluctuations are solely due to or dominated by fluctuations of the flow angle. For \Fvn{2}, i.e., the second-moment flow magnitude fluctuations, this is not the case, as deviations from unity, although small, are observed at all centralities.

\section{Summary}
\label{sec:sum}
Measurements of the \pt-dependent flow vector fluctuations in \PbPb collisions at \snn = 5.02 TeV with the ratio $v_n\{2\}/v_n[2]$ up to $n=4$, and the factorization ratio $r_n$ up to $n=3$ are presented. Deviations of both $v_2\{2\}/v_2[2]$ and $r_2$ from unity suggest the presence of \pt-dependent $\Vn{2}$ fluctuations. The \pt-dependent fluctuations of $\Vn{2}$ reach $\sim 15\%$ in central collisions for $v_2\{2\}/v_2[2]$ at high $\pt$ as well as for $r_2$ when $|p_\mathrm{T}^a-p_\mathrm{T}^t|$ is large. The \pt-dependent flow vector fluctuations of $\Vn{3}$ and $\Vn{4}$ measured via \vncr/\vnsqbr ($n=3,4$) and $r_n$ ($n=3$) are within a few percent. The \vncr/\vnsqbr and $r_n$ results are consistent with previous measurements in \PbPb collisions at $\snnPbOne$~\cite{Acharya:2017ino, Khachatryan:2015oea} but offer significantly better precision. Comparison with the iEBE-VISHNU hydrodynamic model shows that the model with AMPT initial conditions and $\eta/s = 0.08$~\cite{Zhao:2017yhj} or \trento initial conditions and temperature-dependent $\eta/s(T)$ describe flow vector fluctuations well across the presented centrality ranges.

The contributions of flow angle and magnitude fluctuations are separated from the overall flow vector fluctuations with the proposed correlators \Fpsi{n} and \Fvn{n} in various centrality intervals from \cent{0}{50}. Fluctuations of the flow angle and magnitude are observed at higher values of \pt, with the largest fluctuations observed in 0--5\% central collisions. This observation is consistent with the flow vector fluctuations measured with two-particle correlations, exhibiting the largest fluctuations in central collisions. Comparison with hydrodynamic models shows that the AMPT+iEBE-VISHNU with $\eta/s = 0.08$ and \trento+iEBE-VISHNU produce slightly larger flow angle fluctuations compared to the AMPT+iEBE-VISHNU with $\eta/s = 0.20$ in the \cent{0}{5} central collisions. Still, neither of these models can describe the large flow angle fluctuations observed in the data in this centrality interval. For the flow magnitude fluctuations, the hydrodynamic model calculations fail to describe the high $\pt$ deviation from unity in the \cent{0}{5} central collisions. However, the AMPT+iEBE-VISHNU calculation with $\eta/s=0.08$ is closest in describing the data. For non-central collisions, the data are well described by the hydrodynamic calculations.

The flow angle fluctuations dominate all the measured centralities, although significant flow magnitude fluctuations are present, especially in central collisions. The measurements of the flow angle and flow magnitude fluctuations offer an improved understanding of the flow vector fluctuations. They can be used to constrain the initial conditions and transport coefficients of the QGP. Additionally, these measurements suggest that fluctuations of the flow angle are a feature of the system created in heavy-ion collisions and should be incorporated into non-hydrodynamical theoretical models used for comparison with high-\pt flow measurements, where such fluctuations are expected to be more pronounced. 



\newenvironment{acknowledgement}{\relax}{\relax}
\begin{acknowledgement}
\section*{Acknowledgments}
\input{fa_2024-03-04_Opt_C.tex}
\end{acknowledgement}

\bibliographystyle{utphys}
\bibliography{bibliography}
\newpage
\appendix
\section{The ALICE Collaboration}
\label{app:collab}
\input{Alice_Authorlist_2024-03-04_Opt_C.tex}


\end{document}

%% file: fa_2024-03-04_Opt_C.tex

The ALICE Collaboration would like to thank all its engineers and technicians for their invaluable contributions to the construction of the experiment and the CERN accelerator teams for the outstanding performance of the LHC complex.
The ALICE Collaboration gratefully acknowledges the resources and support provided by all Grid centres and the Worldwide LHC Computing Grid (WLCG) collaboration.
The ALICE Collaboration acknowledges the following funding agencies for their support in building and running the ALICE detector:
A. I. Alikhanyan National Science Laboratory (Yerevan Physics Institute) Foundation (ANSL), State Committee of Science and World Federation of Scientists (WFS), Armenia;
Austrian Academy of Sciences, Austrian Science Fund (FWF): [M 2467-N36] and Nationalstiftung f\"{u}r Forschung, Technologie und Entwicklung, Austria;
Ministry of Communications and High Technologies, National Nuclear Research Center, Azerbaijan;
Conselho Nacional de Desenvolvimento Cient\'{\i}fico e Tecnol\'{o}gico (CNPq), Financiadora de Estudos e Projetos (Finep), Funda\c{c}\~{a}o de Amparo \`{a} Pesquisa do Estado de S\~{a}o Paulo (FAPESP) and Universidade Federal do Rio Grande do Sul (UFRGS), Brazil;
Bulgarian Ministry of Education and Science, within the National Roadmap for Research Infrastructures 2020-2027 (object CERN), Bulgaria;
Ministry of Education of China (MOEC) , Ministry of Science \& Technology of China (MSTC) and National Natural Science Foundation of China (NSFC), China;
Ministry of Science and Education and Croatian Science Foundation, Croatia;
Centro de Aplicaciones Tecnol\'{o}gicas y Desarrollo Nuclear (CEADEN), Cubaenerg\'{\i}a, Cuba;
Ministry of Education, Youth and Sports of the Czech Republic, Czech Republic;
The Danish Council for Independent Research | Natural Sciences, the VILLUM FONDEN and Danish National Research Foundation (DNRF), Denmark;
Helsinki Institute of Physics (HIP), Finland;
Commissariat \`{a} l'Energie Atomique (CEA) and Institut National de Physique Nucl\'{e}aire et de Physique des Particules (IN2P3) and Centre National de la Recherche Scientifique (CNRS), France;
Bundesministerium f\"{u}r Bildung und Forschung (BMBF) and GSI Helmholtzzentrum f\"{u}r Schwerionenforschung GmbH, Germany;
General Secretariat for Research and Technology, Ministry of Education, Research and Religions, Greece;
National Research, Development and Innovation Office, Hungary;
Department of Atomic Energy Government of India (DAE), Department of Science and Technology, Government of India (DST), University Grants Commission, Government of India (UGC) and Council of Scientific and Industrial Research (CSIR), India;
National Research and Innovation Agency - BRIN, Indonesia;
Istituto Nazionale di Fisica Nucleare (INFN), Italy;
Japanese Ministry of Education, Culture, Sports, Science and Technology (MEXT) and Japan Society for the Promotion of Science (JSPS) KAKENHI, Japan;
Consejo Nacional de Ciencia (CONACYT) y Tecnolog\'{i}a, through Fondo de Cooperaci\'{o}n Internacional en Ciencia y Tecnolog\'{i}a (FONCICYT) and Direcci\'{o}n General de Asuntos del Personal Academico (DGAPA), Mexico;
Nederlandse Organisatie voor Wetenschappelijk Onderzoek (NWO), Netherlands;
The Research Council of Norway, Norway;
Pontificia Universidad Cat\'{o}lica del Per\'{u}, Peru;
Ministry of Education and Science, National Science Centre and WUT ID-UB, Poland;
Korea Institute of Science and Technology Information and National Research Foundation of Korea (NRF), Republic of Korea;
Ministry of Education and Scientific Research, Institute of Atomic Physics, Ministry of Research and Innovation and Institute of Atomic Physics and Universitatea Nationala de Stiinta si Tehnologie Politehnica Bucuresti, Romania;
Ministry of Education, Science, Research and Sport of the Slovak Republic, Slovakia;
National Research Foundation of South Africa, South Africa;
Swedish Research Council (VR) and Knut \& Alice Wallenberg Foundation (KAW), Sweden;
European Organization for Nuclear Research, Switzerland;
Suranaree University of Technology (SUT), National Science and Technology Development Agency (NSTDA) and National Science, Research and Innovation Fund (NSRF via PMU-B B05F650021), Thailand;
Turkish Energy, Nuclear and Mineral Research Agency (TENMAK), Turkey;
National Academy of  Sciences of Ukraine, Ukraine;
Science and Technology Facilities Council (STFC), United Kingdom;
National Science Foundation of the United States of America (NSF) and United States Department of Energy, Office of Nuclear Physics (DOE NP), United States of America.
In addition, individual groups or members have received support from:
Czech Science Foundation (grant no. 23-07499S), Czech Republic;
European Research Council (grant no. 950692), European Union;
ICSC - Centro Nazionale di Ricerca in High Performance Computing, Big Data and Quantum Computing, European Union - NextGenerationEU;
Academy of Finland (Center of Excellence in Quark Matter) (grant nos. 346327, 346328), Finland.

%% file: Alice_Authorlist_2024-03-04_Opt_C.tex
\begin{flushleft} 
\small

S.~Acharya\,\orcidlink{0000-0002-9213-5329}\,$^{\rm 127}$, 
D.~Adamov\'{a}\,\orcidlink{0000-0002-0504-7428}\,$^{\rm 86}$, 
A.~Agarwal$^{\rm 135}$, 
G.~Aglieri Rinella\,\orcidlink{0000-0002-9611-3696}\,$^{\rm 32}$, 
L.~Aglietta$^{\rm 24}$, 
M.~Agnello\,\orcidlink{0000-0002-0760-5075}\,$^{\rm 29}$, 
N.~Agrawal\,\orcidlink{0000-0003-0348-9836}\,$^{\rm 25}$, 
Z.~Ahammed\,\orcidlink{0000-0001-5241-7412}\,$^{\rm 135}$, 
S.~Ahmad\,\orcidlink{0000-0003-0497-5705}\,$^{\rm 15}$, 
S.U.~Ahn\,\orcidlink{0000-0001-8847-489X}\,$^{\rm 71}$
% ...

\section*{Affiliation Notes}

$^{\rm I}$ Deceased\\
$^{\rm II}$ Also at: Max-Planck-Institut fur Physik, Munich, Germany\\
$^{\rm III}$ Also at: Italian National Agency for New Technologies, Energy and Sustainable Economic Development (ENEA), Bologna, Italy\\
$^{\rm IV}$ Also at: Dipartimento DET del Politecnico di Torino, Turin, Italy\\
$^{\rm V}$ Also at: Yildiz Technical University, Istanbul, T\"{u}rkiye\\
$^{\rm VI}$ Also at: Department of Applied Physics, Aligarh Muslim University, Aligarh, India\\
$^{\rm VII}$ Also at: Institute of Theoretical Physics, University of Wroclaw, Poland\\
$^{\rm VIII}$ Also at: An institution covered by a cooperation agreement with CERN\\

\section*{Collaboration Institutes}

$^{1}$ A.I. Alikhanyan National Science Laboratory (Yerevan Physics Institute) Foundation, Yerevan, Armenia\\
$^{2}$ AGH University of Krakow, Cracow, Poland\\
$^{3}$ Bogolyubov Institute for Theoretical Physics, National Academy of Sciences of Ukraine, Kiev, Ukraine\\
$^{4}$ Bose Institute, Department of Physics  and Centre for Astroparticle Physics and Space Science (CAPSS), Kolkata, India\\
$^{5}$ California Polytechnic State University, San Luis Obispo, California, United States\\
$^{6}$ Central China Normal University, Wuhan, China\\
$^{7}$ Centro de Aplicaciones Tecnol\'{o}gicas y Desarrollo Nuclear (CEADEN), Havana, Cuba\\
$^{8}$ Centro de Investigaci\'{o}n y de Estudios Avanzados (CINVESTAV), Mexico City and M\'{e}rida, Mexico\\
$^{9}$ Chicago State University, Chicago, Illinois, United States\\
$^{10}$ China Institute of Atomic Energy, Beijing, China\\
$^{11}$ China University of Geosciences, Wuhan, China\\
$^{12}$ Chungbuk National University, Cheongju, Republic of Korea\\
$^{13}$ Comenius University Bratislava, Faculty of Mathematics, Physics and Informatics, Bratislava, Slovak Republic\\
$^{14}$ Creighton University, Omaha, Nebraska, United States\\
$^{15}$ Department of Physics, Aligarh Muslim University, Aligarh, India\\
$^{16}$ Department of Physics, Pusan National University, Pusan, Republic of Korea\\
$^{17}$ Department of Physics, Sejong University, Seoul, Republic of Korea\\
$^{18}$ Department of Physics, University of California, Berkeley, California, United States\\
$^{19}$ Department of Physics, University of Oslo, Oslo, Norway\\
$^{20}$ Department of Physics and Technology, University of Bergen, Bergen, Norway\\
$^{21}$ Dipartimento di Fisica, Universit\`{a} di Pavia, Pavia, Italy\\
$^{22}$ Dipartimento di Fisica dell'Universit\`{a} and Sezione INFN, Cagliari, Italy\\
$^{23}$ Dipartimento di Fisica dell'Universit\`{a} and Sezione INFN, Trieste, Italy\\
$^{24}$ Dipartimento di Fisica dell'Universit\`{a} and Sezione INFN, Turin, Italy\\
$^{25}$ Dipartimento di Fisica e Astronomia dell'Universit\`{a} and Sezione INFN, Bologna, Italy\\
$^{26}$ Dipartimento di Fisica e Astronomia dell'Universit\`{a} and Sezione INFN, Catania, Italy\\
$^{27}$ Dipartimento di Fisica e Astronomia dell'Universit\`{a} and Sezione INFN, Padova, Italy\\
$^{28}$ Dipartimento di Fisica `E.R.~Caianiello' dell'Universit\`{a} and Gruppo Collegato INFN, Salerno, Italy\\
$^{29}$ Dipartimento DISAT del Politecnico and Sezione INFN, Turin, Italy\\
$^{30}$ Dipartimento di Scienze MIFT, Universit\`{a} di Messina, Messina, Italy\\
$^{31}$ Dipartimento Interateneo di Fisica `M.~Merlin' and Sezione INFN, Bari, Italy\\
$^{32}$ European Organization for Nuclear Research (CERN), Geneva, Switzerland\\
$^{33}$ Faculty of Electrical Engineering, Mechanical Engineering and Naval Architecture, University of Split, Split, Croatia\\
$^{34}$ Faculty of Engineering and Science, Western Norway University of Applied Sciences, Bergen, Norway\\
$^{35}$ Faculty of Nuclear Sciences and Physical Engineering, Czech Technical University in Prague, Prague, Czech Republic\\
$^{36}$ Faculty of Physics, Sofia University, Sofia, Bulgaria\\
$^{37}$ Faculty of Science, P.J.~\v{S}af\'{a}rik University, Ko\v{s}ice, Slovak Republic\\
$^{38}$ Frankfurt Institute for Advanced Studies, Johann Wolfgang Goethe-Universit\"{a}t Frankfurt, Frankfurt, Germany\\
$^{39}$ Fudan University, Shanghai, China\\
$^{40}$ Gangneung-Wonju National University, Gangneung, Republic of Korea\\
$^{41}$ Gauhati University, Department of Physics, Guwahati, India\\
$^{42}$ Helmholtz-Institut f\"{u}r Strahlen- und Kernphysik, Rheinische Friedrich-Wilhelms-Universit\"{a}t Bonn, Bonn, Germany\\
$^{43}$ Helsinki Institute of Physics (HIP), Helsinki, Finland\\
$^{44}$ High Energy Physics Group,  Universidad Aut\'{o}noma de Puebla, Puebla, Mexico\\
$^{45}$ Horia Hulubei National Institute of Physics and Nuclear Engineering, Bucharest, Romania\\
$^{46}$ HUN-REN Wigner Research Centre for Physics, Budapest, Hungary\\
$^{47}$ Indian Institute of Technology Bombay (IIT), Mumbai, India\\
$^{48}$ Indian Institute of Technology Indore, Indore, India\\
$^{49}$ INFN, Laboratori Nazionali di Frascati, Frascati, Italy\\
$^{50}$ INFN, Sezione di Bari, Bari, Italy\\
$^{51}$ INFN, Sezione di Bologna, Bologna, Italy\\
$^{52}$ INFN, Sezione di Cagliari, Cagliari, Italy\\
$^{53}$ INFN, Sezione di Catania, Catania, Italy\\
$^{54}$ INFN, Sezione di Padova, Padova, Italy\\
$^{55}$ INFN, Sezione di Pavia, Pavia, Italy\\
$^{56}$ INFN, Sezione di Torino, Turin, Italy\\
$^{57}$ INFN, Sezione di Trieste, Trieste, Italy\\
$^{58}$ Inha University, Incheon, Republic of Korea\\
$^{59}$ Institute for Gravitational and Subatomic Physics (GRASP), Utrecht University/Nikhef, Utrecht, Netherlands\\
$^{60}$ Institute of Experimental Physics, Slovak Academy of Sciences, Ko\v{s}ice, Slovak Republic\\
$^{61}$ Institute of Physics, Homi Bhabha National Institute, Bhubaneswar, India\\
$^{62}$ Institute of Physics of the Czech Academy of Sciences, Prague, Czech Republic\\
$^{63}$ Institute of Space Science (ISS), Bucharest, Romania\\
$^{64}$ Institut f\"{u}r Kernphysik, Johann Wolfgang Goethe-Universit\"{a}t Frankfurt, Frankfurt, Germany\\
$^{65}$ Instituto de Ciencias Nucleares, Universidad Nacional Aut\'{o}noma de M\'{e}xico, Mexico City, Mexico\\
$^{66}$ Instituto de F\'{i}sica, Universidade Federal do Rio Grande do Sul (UFRGS), Porto Alegre, Brazil\\
$^{67}$ Instituto de F\'{\i}sica, Universidad Nacional Aut\'{o}noma de M\'{e}xico, Mexico City, Mexico\\
$^{68}$ iThemba LABS, National Research Foundation, Somerset West, South Africa\\
$^{69}$ Jeonbuk National University, Jeonju, Republic of Korea\\
$^{70}$ Johann-Wolfgang-Goethe Universit\"{a}t Frankfurt Institut f\"{u}r Informatik, Fachbereich Informatik und Mathematik, Frankfurt, Germany\\
$^{71}$ Korea Institute of Science and Technology Information, Daejeon, Republic of Korea\\
$^{72}$ KTO Karatay University, Konya, Turkey\\
$^{73}$ Laboratoire de Physique Subatomique et de Cosmologie, Universit\'{e} Grenoble-Alpes, CNRS-IN2P3, Grenoble, France\\
$^{74}$ Lawrence Berkeley National Laboratory, Berkeley, California, United States\\
$^{75}$ Lund University Department of Physics, Division of Particle Physics, Lund, Sweden\\
$^{76}$ Nagasaki Institute of Applied Science, Nagasaki, Japan\\
$^{77}$ Nara Women{'}s University (NWU), Nara, Japan\\
$^{78}$ National and Kapodistrian University of Athens, School of Science, Department of Physics , Athens, Greece\\
$^{79}$ National Centre for Nuclear Research, Warsaw, Poland\\
$^{80}$ National Institute of Science Education and Research, Homi Bhabha National Institute, Jatni, India\\
$^{81}$ National Nuclear Research Center, Baku, Azerbaijan\\
$^{82}$ National Research and Innovation Agency - BRIN, Jakarta, Indonesia\\
$^{83}$ Niels Bohr Institute, University of Copenhagen, Copenhagen, Denmark\\
$^{84}$ Nikhef, National institute for subatomic physics, Amsterdam, Netherlands\\
$^{85}$ Nuclear Physics Group, STFC Daresbury Laboratory, Daresbury, United Kingdom\\
$^{86}$ Nuclear Physics Institute of the Czech Academy of Sciences, Husinec-\v{R}e\v{z}, Czech Republic\\
$^{87}$ Oak Ridge National Laboratory, Oak Ridge, Tennessee, United States\\
$^{88}$ Ohio State University, Columbus, Ohio, United States\\
$^{89}$ Physics department, Faculty of science, University of Zagreb, Zagreb, Croatia\\
$^{90}$ Physics Department, Panjab University, Chandigarh, India\\
$^{91}$ Physics Department, University of Jammu, Jammu, India\\
$^{92}$ Physics Program and International Institute for Sustainability with Knotted Chiral Meta Matter (SKCM2), Hiroshima University, Hiroshima, Japan\\
$^{93}$ Physikalisches Institut, Eberhard-Karls-Universit\"{a}t T\"{u}bingen, T\"{u}bingen, Germany\\
$^{94}$ Physikalisches Institut, Ruprecht-Karls-Universit\"{a}t Heidelberg, Heidelberg, Germany\\
$^{95}$ Physik Department, Technische Universit\"{a}t M\"{u}nchen, Munich, Germany\\
$^{96}$ Politecnico di Bari and Sezione INFN, Bari, Italy\\
$^{97}$ Research Division and ExtreMe Matter Institute EMMI, GSI Helmholtzzentrum f\"ur Schwerionenforschung GmbH, Darmstadt, Germany\\
$^{98}$ Saga University, Saga, Japan\\
$^{99}$ Saha Institute of Nuclear Physics, Homi Bhabha National Institute, Kolkata, India\\
$^{100}$ School of Physics and Astronomy, University of Birmingham, Birmingham, United Kingdom\\
$^{101}$ Secci\'{o}n F\'{\i}sica, Departamento de Ciencias, Pontificia Universidad Cat\'{o}lica del Per\'{u}, Lima, Peru\\
$^{102}$ Stefan Meyer Institut f\"{u}r Subatomare Physik (SMI), Vienna, Austria\\
$^{103}$ SUBATECH, IMT Atlantique, Nantes Universit\'{e}, CNRS-IN2P3, Nantes, France\\
$^{104}$ Sungkyunkwan University, Suwon City, Republic of Korea\\
$^{105}$ Suranaree University of Technology, Nakhon Ratchasima, Thailand\\
$^{106}$ Technical University of Ko\v{s}ice, Ko\v{s}ice, Slovak Republic\\
$^{107}$ The Henryk Niewodniczanski Institute of Nuclear Physics, Polish Academy of Sciences, Cracow, Poland\\
$^{108}$ The University of Texas at Austin, Austin, Texas, United States\\
$^{109}$ Universidad Aut\'{o}noma de Sinaloa, Culiac\'{a}n, Mexico\\
$^{110}$ Universidade de S\~{a}o Paulo (USP), S\~{a}o Paulo, Brazil\\
$^{111}$ Universidade Estadual de Campinas (UNICAMP), Campinas, Brazil\\
$^{112}$ Universidade Federal do ABC, Santo Andre, Brazil\\
$^{113}$ Universitatea Nationala de Stiinta si Tehnologie Politehnica Bucuresti, Bucharest, Romania\\
$^{114}$ University of Cape Town, Cape Town, South Africa\\
$^{115}$ University of Derby, Derby, United Kingdom\\
$^{116}$ University of Houston, Houston, Texas, United States\\
$^{117}$ University of Jyv\"{a}skyl\"{a}, Jyv\"{a}skyl\"{a}, Finland\\
$^{118}$ University of Kansas, Lawrence, Kansas, United States\\
$^{119}$ University of Liverpool, Liverpool, United Kingdom\\
$^{120}$ University of Science and Technology of China, Hefei, China\\
$^{121}$ University of South-Eastern Norway, Kongsberg, Norway\\
$^{122}$ University of Tennessee, Knoxville, Tennessee, United States\\
$^{123}$ University of the Witwatersrand, Johannesburg, South Africa\\
$^{124}$ University of Tokyo, Tokyo, Japan\\
$^{125}$ University of Tsukuba, Tsukuba, Japan\\
$^{126}$ Universit\"{a}t M\"{u}nster, Institut f\"{u}r Kernphysik, M\"{u}nster, Germany\\
$^{127}$ Universit\'{e} Clermont Auvergne, CNRS/IN2P3, LPC, Clermont-Ferrand, France\\
$^{128}$ Universit\'{e} de Lyon, CNRS/IN2P3, Institut de Physique des 2 Infinis de Lyon, Lyon, France\\
$^{129}$ Universit\'{e} de Strasbourg, CNRS, IPHC UMR 7178, F-67000 Strasbourg, France, Strasbourg, France\\
$^{130}$ Universit\'{e} Paris-Saclay, Centre d'Etudes de Saclay (CEA), IRFU, D\'{e}partment de Physique Nucl\'{e}aire (DPhN), Saclay, France\\
$^{131}$ Universit\'{e}  Paris-Saclay, CNRS/IN2P3, IJCLab, Orsay, France\\
$^{132}$ Universit\`{a} degli Studi di Foggia, Foggia, Italy\\
$^{133}$ Universit\`{a} del Piemonte Orientale, Vercelli, Italy\\
$^{134}$ Universit\`{a} di Brescia, Brescia, Italy\\
$^{135}$ Variable Energy Cyclotron Centre, Homi Bhabha National Institute, Kolkata, India\\
$^{136}$ Warsaw University of Technology, Warsaw, Poland\\
$^{137}$ Wayne State University, Detroit, Michigan, United States\\
$^{138}$ Yale University, New Haven, Connecticut, United States\\
$^{139}$ Yonsei University, Seoul, Republic of Korea\\
$^{140}$  Zentrum  f\"{u}r Technologie und Transfer (ZTT), Worms, Germany\\
$^{141}$ Affiliated with an institute covered by a cooperation agreement with CERN\\
$^{142}$ Affiliated with an international laboratory covered by a cooperation agreement with CERN.\\

\end{flushleft} 